\documentclass[twocolumn,showkeys]{revtex4}
\usepackage[utf8]{inputenc}
\usepackage{graphicx}

\begin{document}

\title{Nonlinear Coupling of Electromagnetic and Spin-Electron-Acoustic Waves in Spin-polarized
Degenerate Relativistic Astrophysical Plasma}

\author{Pavel A. Andreev}
\email{andreevpa@physics.msu.ru}
\affiliation{Department of General Physics, Faculty of physics, Lomonosov Moscow State University, Moscow, Russian Federation, 119991.}

\date{\today}

\begin{abstract}
Propagation of the finite amplitude electromagnetic wave through the partially spin-polarized degenerate plasmas
leads to the instability.
The instability happens at the interaction of the electromagnetic wave
with the small frequency longitudinal spin-electron-acoustic waves.
Strongest instability happens in the high density degenerate plasmas with the Fermi momentum close to $m_{e}c$,
where $m_{e}$ is the mass of electron,
and $c$ is the speed of light.
The increase of the increment of instability with the growth of the spin polarization of plasmas is found.
\end{abstract}



\maketitle





\section{Introduction}

We consider the propagation of the large amplitude high frequency electromagnetic radiation
through the high density plasmas surrounding the compact astrophysical objects, such as white dwarfs or neutron stars.
Mostly these objects generate strong magnetic fields
which modifies the trajectory of macroscopic flows or individual charged particles.
It also strongly changes the spectrum of longitudinal and electromagnetic waves propagating in the highly magnetized plasmas,
creating, for instance, the longitudinally-transverse waves.
The magnetic fields create the spin-polarization of electrons as well,
which can be macroscopically hidden by the diamagnetic effects of moving charges.
Propagation of the electromagnetic waves in plasmas leads to the spin polarization of medium via the relativistic effects
of the spin-torque \cite{Brodin PRL 10}, the spin-orbit interaction \cite{Mochizuki APL 18},
or presence small amplitude low frequency electromagnetic radiation absorbed by electrons with further spin flipping.

The white dwarfs are the least compact of the compact astrophysical objects.
But their density is enough to reach high density degenerate plasmas with the Fermi momentum close to $m_{e}c$,
where $m_{e}$ is the mass of electron,
and $c$ is the speed of light.
It is possible that the observed radiation
which comes from the compact astrophysical objects can give information
from nonlinear interactions,
particularly about interaction of the high-frequency electromagnetic radiation with the matter waves existing in plasmas
\cite{Akbari PoP 13}, \cite{Berezhiani PP 15}, \cite{Shatashvili PoP 20}.
There are examples of plasmas composed of two species of electrons
which demonstrate scattering of the large amplitude electromagnetic waves on the sound waves,
particularly the decay instability and the modulation instability are found in Ref. \cite{Shatashvili PoP 20}.

Plasmas can exist in extreme conditions, near the compact objects with magnetic field close to critical value,
where the particle-antiparticle pairs can occur
\cite{Uzdensky RPP review 14}.
However, we do not consider these kinds of objects here.


The partially spin polarized degenerate plasmas show existence of the spin-electron-acoustic waves
\cite{Andreev PRE 15}, \cite{Andreev EPL 16}, \cite{Andreev PRE 16}, \cite{Andreev PP 16 SSE kin},
\cite{Ryan PRB 91}, \cite{Agarwal PRB 14}.
Particularly, the relativistic spin-electron-acoustic waves are studied in high-density degenerate spin-polarized plasmas
using the separate-spin-evolution relativistic hydrodynamic model with the average reverse gamma factor evolution \cite{Andreev 2112}.
This hydrodynamic model is partially based on the quantum hydrodynamic models
\cite{Andreev PRE 15}, \cite{Andreev Ch 21}, \cite{Andreev PoF 21}, \cite{Andreev JPP 21}.
Variety of instabilities are studied in nonrelativistic degenerate plasmas,
particularly effects appearing due to the separate spin evolution of electrons with different spin projections
\cite{Iqbal IEEE 22}, \cite{Iqbal Shahid PP 17}, \cite{Iqbal PLA 18},
\cite{Iqbal PP 18 UHW}, \cite{Iqbal PP 18 LI}.

In this paper, we combine the knowledge about the relativistic spin-electron-acoustic wave with the possibility nonlinear instabilities
existing in degenerate plasmas surrounding the white dwarfs in order to find new regimes for the unstable plasma behavior.

This paper is organized as follows.
In Sec. II the separate spin evolution relativistic hydrodynamic model with the average reverse gamma factor evolution
is presented for the systems of degenerate partially spin polarized electrons.
In Sec. III approximate wave equations for the coupled transverse and longitudinal waves are derived.
In Sec. IV the analysis of stability of the spin-electron-acoustic waves under pumping of electromagnetic wave is analyzed.
In Sec. IV a brief summary of obtained results is presented.



\section{Two fluid relativistic hydrodynamic model with the average reverse gamma factor evolution}

In order to study waves in the high-density degenerate spin-polarized plasmas
we apply the separate spin evolution relativistic hydrodynamic model with the average reverse gamma factor evolution,
which is developed in Ref. \cite{Andreev 2112}
as the generalization of Refs. \cite{Andreev 2021 05}, \cite{Andreev 2021 06}, \cite{Andreev 2021 08}, \cite{Andreev 2021 09}, \cite{Andreev 2021 11}.
This model is composed of four following equations for each species of charged particles
which are obtained in the mean-field approximation.
One is the continuity equation
\begin{equation}\label{RHDSEAW2022NLC cont via v} \partial_{t}n_{s}+\nabla\cdot(n_{s}\textbf{v}_{s})=0.\end{equation}
Second equation is a form of relativistic Euler equation derived for the flux of particles
and showing the evolution of the velocity field
$$n_{s}\partial_{t}\textbf{v}_{s}+n_{s}(\textbf{v}_{s}\cdot\nabla)\textbf{v}_{s}
+\frac{1}{m_{s}}\nabla\tilde{p}_{s}$$
$$=\frac{q_{s}}{m_{s}}\biggl(\Gamma_{s} -\frac{\tilde{t}_{s}}{c^{2}}\biggr)\textbf{E}
+\frac{q_{s}}{m_{s}c}[(\Gamma_{s} \textbf{v}_{s}+\textbf{t}_{s})\times\textbf{B}]$$
\begin{equation}\label{RHDSEAW2022NLC Euler for v}
-\frac{q_{s}}{m_{s}c^{2}}
\biggl(\Gamma_{s} \textbf{v}_{s} (\textbf{v}_{s}\cdot \textbf{E})
+\textbf{v}_{s} (\textbf{t}_{s}\cdot\textbf{E})+\textbf{t}_{s} (\textbf{v}_{s}\cdot\textbf{E})\biggr), \end{equation}
where
$m_{s}$ and $q_{s}$ are the mass and charge of particle of $s$ species,
$c$ is the speed of light,
tensor $p^{ab}_{s}=\tilde{p}_{s}\delta^{ab}$
is the flux of the velocities for electrons with fixed spin projection,
tensor $t^{ab}_{s}=\tilde{t}_{s}\delta^{ab}$
is the flux of the average reverse gamma-factor for spin-s electrons,
$\varepsilon^{abc}$ is the three-dimensional Levi-Civita symbol,
$\delta^{ab}$ is the three-dimensional Kronecker symbol.
Moreover, we work in the Minkovskii space, hence the metric tensor has diagonal form
with the following sings $g^{\alpha\beta}=\{-1, +1, +1, +1\}$.
We mostly use the three dimensional notations,
therefore, we can change covariant and contrvariant indexes for the three-vector indexes: $v^{b}_{s}=v_{b,s}$.
The Latin indexes like $a$, $b$, $c$ etc describe the three-vectors,
while the Greek indexes are deposited for the four-vector notations.
We also have Latin index $s$ which refers to the species or subspecies of electrons with different spin projections.
However, the indexes related to coordinates are chosen from the beginning of the alphabet,
while other indexes are chosen in accordance with their physical meaning.
The model under presentation and this Euler equation includes no effects related to change of the spin projections of electrons on the chosen direction.

The third equation is for the evolution for the average reverse relativistic gamma factor
or the hydrodynamic Gamma function
$$\partial_{t}\Gamma_{s}+\nabla(\Gamma_{s} \textbf{v}_{s}+\textbf{t}_{s})$$
\begin{equation}\label{RHDSEAW2022NLC eq for Gamma}
=-\frac{q_{s}}{m_{s}c^{2}}n(\textbf{v}_{s}\cdot\textbf{E})
\biggl(1-\frac{1}{c^{2}}\biggl(\textbf{v}_{s}^{2}+\frac{5\tilde{p}_{s}}{n_{s}}\biggr)\biggr).\end{equation}
The fourth and final material equation is the equation for the evolution of the flux of reverse gamma factor
$$(\partial_{t}+\textbf{v}_{s}\cdot\nabla)\textbf{t}_{s}^{a}
+\nabla ^{a}\tilde{t}_{s}
+(\textbf{t}_{s}\cdot\nabla) \textbf{v}_{s}^{a}+\textbf{t}_{s}^{a} (\nabla\cdot \textbf{v}_{s})$$
$$+\Gamma_{s}(\partial_{t}+\textbf{v}_{s}\cdot\nabla)\textbf{v}_{s}^{a}
=\frac{q_{s}}{m_{s}}n_{s}\textbf{E}^{a}\biggl[1-\frac{\textbf{v}_{s}^{2}}{c^{2}}-\frac{3\tilde{p}_{s}}{n_{s}c^{2}}\biggr]$$
$$+\frac{q_{s}}{m_{s}c}[n_{s}\textbf{v}_{s}\times \textbf{B}]
\biggl[1-\frac{\textbf{v}_{s}^{2}}{c^{2}}-\frac{5\tilde{p}_{s}}{n_{s}c^{2}}\biggr]
-\frac{2q_{s}}{m_{s}c^{2}}\Biggl[\textbf{E}^{a}\tilde{p}_{s}\biggl(1-\frac{\textbf{v}_{s}^{2}}{c^2}\biggr)$$
\begin{equation}\label{RHDSEAW2022NLC eq for t a}
+n_{s}\textbf{v}_{s}^{a}(\textbf{v}_{s}\cdot\textbf{E})\biggl(1-\frac{\textbf{v}_{s}^{2}}{c^{2}}-\frac{9\tilde{p}_{s}}{n_{s}c^{2}}\biggr)
-\frac{5M_{s0}}{3c^{2}} \textbf{E}^{a}\Biggr].\end{equation}
Some functions appearing in this set of equations are discussed below together with the necessary equations of state.



The Maxwell equations are used to couple interspecies and inspecies electromagnetic interaction
$\nabla \cdot\textbf{B}=0$,
\begin{equation}\label{RHDSEAW2022NLC rot E} \nabla\times \textbf{E}=-\frac{1}{c}\partial_{t}\textbf{B},\end{equation}
\begin{equation}\label{RHDSEAW2022NLC div E with time} \nabla \cdot\textbf{E}=
4\pi(q_{e}n_{e\uparrow}+q_{e\downarrow}n_{e}+q_{i}n_{i}),
\end{equation}
and
\begin{equation}\label{RHDSEAW2022NLC rot B with time}
\nabla\times \textbf{B}=\frac{1}{c}\partial_{t}\textbf{E}+
\frac{4\pi q_{e}}{c}n_{e\uparrow}\textbf{v}_{e\uparrow}
+\frac{4\pi q_{e}}{c}n_{e\downarrow}\textbf{v}_{e\downarrow}
+\frac{4\pi q_{i}}{c}n_{i}\textbf{v}_{i}.
\end{equation}
In this paper we consider the ions as the motionless background,
hence below we have $\textbf{v}_{i}=0$.

Presented hydrodynamic equations do not contain information about evolution of the spin density.
However, there are nonrelativistic quantum hydrodynamic and kinetic models,
where this effect included \cite{Andreev PRE 15}, \cite{Andreev PoF 21},
\cite{Andreev PoP kinetics 17 a}, \cite{Andreev PoP non-triv kinetics 17}.
Moreover, some relativistic spin effects in plasmas are modeled via hydrodynamics in Ref. \cite{Dodin PRA 15 First-principle}.
The variety of other models describing the relativistic plasmas can be found in literature
\cite{Mahajan PoP 16}, \cite{Brunetti MNRAS 04}, \cite{Munoz EPS 06}, \cite{Mikaberidze PLA 15},
\cite{Asenjo 19}, \cite{Liu PPCF 21}, \cite{Ekman PRE 17}, \cite{Shatashvili ASS 97},
\cite{Heyvaerts AA 12}, \cite{Comisso 19}.

\subsection{Equations of state for spin-up and spin-down electrons}

We follow the results of Ref. \cite{Andreev 2112} to present the equations of state,
which are necessary to get the closed set of the relativistic hydrodynamic equations.
Let us repeat the method of derivation of equations of state.
We consider the high density degenerate electron gas.
The Fermi velocity is obtained for the relativistic regime
$v_{Fs}=p_{Fs}/\sqrt{1+p_{Fs}^{2}/m_{s}^{2}c^{2}}m_{s}$,
where $p_{Fs}=(6\pi^{2}n_{s})^{1/3}\hbar$.

Systems of degenerate fermions with the fixed spin projection are described within the Fermi-Dirac distribution,
which simplifies to the Fermi step distribution
for the zero-temperature limit
\begin{equation}\label{RHDSEAW2022NLC Fermi step} f_{s0}=\Biggl\{\begin{array}{c}
                                                               \frac{1}{(2\pi\hbar)^{3}} \\
                                                               0
                                                             \end{array}
\textrm{for}
\begin{array}{c}
                                                               p\leq p_{Fs} \\
                                                               p> p_{Fs}
                                                             \end{array}
\end{equation}
The concentration of $s$-species can be expressed via the distribution function
\begin{equation}\label{RHDSEAW2022NLC concentr via f} n_{s}=\int f_{s0} d^{3}p. \end{equation}
where
$\textbf{p}=m_{s}\textbf{v}/\sqrt{1-\textbf{v}^{2}/c^{2}}$.

Here we ready to present the flux of the current of particles via the distribution function
\begin{equation}\label{RHDSEAW2022NLC p via f} p_{s}^{ab}=\int v^{a}v^{b} f_{s0} d^{3}p. \end{equation}
We use distribution function (\ref{RHDSEAW2022NLC Fermi step}) to calculate
the equation of state $p_{s}^{ab}=\tilde{p}_{s} \delta^{ab}$:
\begin{equation}\label{RHDSEAW2022NLC p rel eq of state}
\tilde{p}_{s}=\frac{m_{s}^{3}c^{5}}{6\pi^{2}\hbar^{3}}\biggl[\frac{1}{3}\xi_{s}^{3}-\xi_{s}+\arctan\xi_{s}\biggr],
\end{equation}
where
$\xi_{s}\equiv p_{Fs}/mc$.

Next, we derive the flux of the current of the average reverse gamma factor via the distribution function
\begin{equation}\label{RHDSEAW2022NLC t via f} t_{s}^{ab}=\int \biggl(\frac{v^{a}v^{b}}{\gamma}\biggr) f_{s0} d^{3}p, \end{equation}
with the following result $t^{ab}_{s}=\tilde{t}_{s} \delta^{ab}$,
and
\begin{equation}\label{RHDSEAW2022NLC t rel eq of state}
\tilde{t}_{s}=\frac{m_{s}^{3}c^{5}}{12\pi^{2}\hbar^{3}} \biggl[ \xi_{s}\sqrt{\xi_{s}^{2}+1}+\frac{2\xi_{s}}{\sqrt{\xi_{s}^{2}+1}} -3Arsinh\xi_{s}\biggr],
\end{equation}
where
$Arsinh\xi=ln\mid \xi+\sqrt{\xi^{2}+1}\mid$,
and
$sinh(Arsinh\xi)=\xi$.

The fourth rank tensor $M_{s}^{abcd}$ is
\begin{equation}\label{RHDSEAW2022NLC M via f} M_{s}^{abcd}=\int v^{a}v^{b} v^{c}v^{d} f_{s0} d^{3}p. \end{equation}
Expression (\ref{RHDSEAW2022NLC M rel eq of state}) leads to
$M_{s}^{abcd}=(M_{s0}/3)(\delta^{ab}\delta^{cd}+\delta^{ac}\delta^{bd}+\delta^{ad}\delta^{bc})$,
where
\begin{equation}\label{RHDSEAW2022NLC M rel eq of state}
M_{s0}=\frac{m_{s}^{3}c^{7}}{60\pi^{2}\hbar^{3}} \biggl[ 2\xi_{s}(\xi_{s}^{2} -6) -\frac{3\xi_{s}}{\xi_{s}^{2}+1} +15\arctan\xi_{s} \biggr].
\end{equation}
The fourth rank tensor $M^{abcd}_{s}$ enters equation
(\ref{RHDSEAW2022NLC eq for t a}) via its partial trace $M^{abcc}_{s}=M_{s,c}^{cab}$.
We have the following nonzero elements of this tensor: $M_{s}^{xxxx}=M_{s}^{yyyy}=M_{s}^{zzzz}=M_{s0}$
and $M_{s}^{xxyy}=M_{s}^{xxzz}=M_{s0}/3$.
The partial trace $M_{s,c}^{cab}$ has the following presentation via elements of tensor $M_{s0}$:
$M_{s,c}^{cab}=(5M_{s0}/3)\delta^{ab}$.


We also find the equilibrium expression for function $\Gamma_{0s}$:
\begin{equation}\label{RHDSEAW2022NLC Gamma via f} \Gamma_{s}=\int \frac{1}{\gamma} f_{s0} d^{3}p, \end{equation}
with
\begin{equation}\label{RHDSEAW2022NLC Gamma rel eq of state}
\Gamma_{0s}= \frac{m_{s}^{3}c^{3}}{4\pi^{2}\hbar^{3}} \biggl[ \xi_{s}\sqrt{\xi_{s}^{2}+1} -Arsinh\xi_{s}\biggr].
\end{equation}

\section{Wave equations for the longitudinal spin-electron-acoustic waves in presence of the finite amplitude electromagnetic wave}

Presence of the strong magnetic field leads to change of the behavior of waves and plasmas.
We neglect this effect considering white dwarfs with relatively small magnetic field.
However, presented analysis can be considered as the rough estimation of possible instabilities
which may occur in the plasmas surrounding the neutron stars.

We rewrite the electromagnetic field in terms of the scalar $\varphi$ and vector $\textbf{A}$ potentials
$\textbf{E}=-\nabla\varphi-\frac{1}{c}\partial_{t}\textbf{A}$,
and
$\textbf{B}=\nabla\times\textbf{A}$.
Moreover, we use the Coulomb Gauge $\nabla\cdot \textbf{A}=0$.
Therefore, the Maxwell equations have the following form

\begin{equation}\label{RHDSEAW2022NLC eq for A gen}
\partial_{t}^{2}\textbf{A}-c^{2}\triangle \textbf{A} +c\partial_{t}\nabla\varphi
-4\pi c q_{e}(n_{eu}\textbf{v}_{eu}+n_{ed}\textbf{v}_{ed})=0, \end{equation}
and
\begin{equation}\label{RHDSEAW2022NLC eq for varphi gen} \triangle\varphi=-4\pi (q_{e}n_{eu}+q_{e}n_{ed}+q_{i}n_{i}). \end{equation}

We assume that all hydrodynamic functions depend on coordinate $z$ which is the direction of the electromagnetic wave propagatoion
and time $t$.
For instance, the concentration is $n=n(z,t)$.
Moreover, the vector potential of the electromagnetic field has the following structure
\begin{equation}\label{RHDSEAW2022NLC A structure}
\textbf{A}_{\perp}=\frac{1}{2}\biggl[(\textbf{e}_{x}+\imath \textbf{e}_{y})A(z,t)e^{\imath k_{0}z-\imath\omega_{0} t}+c.c.,\biggr]
\end{equation}
where
$k_{0}$ and
$\omega_{0}$ are the wave vector and the frequency of the electromagnetic wave propagating through the plasmas,
and
$c.c.$ is the complex conjugation.

As the result of suggested structure of functions
we find
$A_{z}=\textbf{e}_{z}\cdot \textbf{A}=0$ from the Coulomb Gauge,
$E_{z}=-\partial_{z}\varphi$,
$\textbf{E}_{\perp}=-\frac{1}{c}\partial_{t}\textbf{A}_{\perp}$,
$\textbf{B}=-\partial_{z}A_{y}\textbf{e}_{x}+\partial_{z}A_{x}\textbf{e}_{y}$.

For the suggested structure of functions
we find the following equation for the transverse part of the vector potential
\begin{equation}\label{RHDSEAW2022NLC eq for A z perp}
\partial_{t}^{2}\textbf{A}_{\perp}-c^{2}\partial_{z}^{2} \textbf{A}_{\perp}
-4\pi c q_{e}(n_{eu}\textbf{v}_{eu\perp}+n_{ed}\textbf{v}_{ed\perp})=0, \end{equation}


We find simplification of equation (\ref{RHDSEAW2022NLC Euler for v}) for the longitudinal perturbations:
$$n_{s}\partial_{t}v_{sz}+n_{s}v_{sz}\partial_{z}v_{sz}+\partial_{z}p_{s}
=-\frac{q_{e}}{m_{e}}\biggl(\Gamma_{s}-\frac{1}{c^{2}}\tilde{t}_{s}\biggr)\partial_{z}\varphi$$
\begin{equation}\label{RHDSEAW2022NLC Euler Long z 1}
+\frac{q_{e}}{m_{e}c}[(\Gamma_{s}v_{sx}+t_{sx})\partial_{z}A_{x}+(\Gamma_{s}v_{sy}+t_{sy})\partial_{z}A_{y}]. \end{equation}

The transverse motion of the electrons is described by the following equation
$$n_{s}\partial_{t}\textbf{v}_{s\perp}+n_{s}v_{sz}\partial_{z}\textbf{v}_{s\perp}
+\frac{q_{e}}{m_{e}c}\biggl(\Gamma_{s}-\frac{1}{c^{2}}\tilde{t}_{s}\biggr)\partial_{t}\textbf{A}_{\perp}$$
\begin{equation}\label{RHDSEAW2022NLC Euler Tr 1}
+\frac{q_{e}}{m_{e}c}(\Gamma_{s}v_{sz}+t_{sz})\partial_{z}\textbf{A}_{\perp}=0 \end{equation}

Equation (\ref{RHDSEAW2022NLC Euler Tr 1}) can be represented in the following form
$$\partial_{t}(n_{s}\textbf{v}_{s\perp})+\partial_{z}(n_{s}v_{sz}\textbf{v}_{s\perp})
+\frac{q_{e}}{m_{e}c}\biggl(\Gamma_{s}-\frac{1}{c^{2}}\tilde{t}_{s}\biggr)\partial_{t}\textbf{A}_{\perp}$$
\begin{equation}\label{RHDSEAW2022NLC Euler Tr 2}
+\frac{q_{e}}{m_{e}c}(\Gamma_{s}v_{sz}+t_{sz})\partial_{z}\textbf{A}_{\perp}=0 \end{equation}
using the continuity equation
\begin{equation}\label{RHDSEAW2022NLC cont z} \partial_{t}n_{s}+\partial_{z}(n_{s}v_{sz})=0.\end{equation}

Next, we try to put coefficients in front of
$\partial_{t}\textbf{A}_{\perp}$ and $\partial_{z}\textbf{A}_{\perp}$
in equation (\ref{RHDSEAW2022NLC Euler Tr 2}) under the derivatives.
To do this step we need to consider equation for the average reverse gamma factor evolution (\ref{RHDSEAW2022NLC eq for Gamma}) in more details.
As we stated above,
we assume that
we consider relatively small intensity electromagnetic waves and nonrelativistic flows.
Hence, the right-hand side of equation (\ref{RHDSEAW2022NLC eq for Gamma}) can be dropped
and this equation reappears in the following form
\begin{equation}\label{RHDSEAW2022NLC Gamma z simpl}
\partial_{t}\Gamma_{s}+\partial_{z}(\Gamma_{s}v_{sz}+t_{sz})=0.\end{equation}

Equation (\ref{RHDSEAW2022NLC Gamma z simpl}) allows to give the following representation of
equation (\ref{RHDSEAW2022NLC Euler Tr 2})
$$\partial_{t}(n_{s}\textbf{v}_{s\perp})+\partial_{z}(n_{s}v_{sz}\textbf{v}_{s\perp})
+\frac{q_{e}}{m_{e}c}\partial_{t}\biggl[\biggl(\Gamma_{s}-\frac{1}{c^{2}}\tilde{t}_{s}\biggr)\textbf{A}_{\perp}\biggr]$$
\begin{equation}\label{RHDSEAW2022NLC Euler Tr 3}
+\frac{q_{e}}{m_{e}c}\partial_{z}[(\Gamma_{s}v_{sz}+t_{sz})\textbf{A}_{\perp}]
+\frac{1}{c^{2}}\frac{q_{e}}{m_{e}c}\textbf{A}_{\perp}\partial_{t}\tilde{t}_{s}=0. \end{equation}
So, we focus on the last term in the found equation:
$\frac{q_{e}}{m_{e}c^{3}}\textbf{A}_{\perp}\partial_{t}\tilde{t}_{s}$
$=\frac{q_{e}}{m_{e}c^{3}}\textbf{A}_{\perp}\frac{\delta \tilde{t}_{s}}{\delta n_{s}}\partial_{t}n_{s}$
$=-\frac{q_{e}}{m_{e}c^{3}}\textbf{A}_{\perp}\frac{\delta \tilde{t}_{s}}{\delta n_{s}}\partial_{z}(n_{s}v_{sz})$.
Here we have $v_{sz}\textbf{A}_{\perp}/c^{2}\sim \textbf{v}^{2}/c^{2}\ll 1$.
It shows that we can neglect the last term in equation (\ref{RHDSEAW2022NLC Euler Tr 3})
and find the following structure
$$\partial_{t}\biggl[n_{s}\textbf{v}_{s\perp}
+\frac{q_{e}}{m_{e}c}\biggl(\Gamma_{s}-\frac{1}{c^{2}}\tilde{t}_{s}\biggr)\textbf{A}_{\perp}\biggr]$$
\begin{equation}\label{RHDSEAW2022NLC Euler Tr 4}
+\partial_{z}\biggl[n_{s}v_{sz}\textbf{v}_{s\perp}
+\frac{q_{e}}{m_{e}c}(\Gamma_{s}v_{sz}+t_{sz})\textbf{A}_{\perp}\biggr]=0. \end{equation}

Equation (\ref{RHDSEAW2022NLC Euler Tr 4}) shows the conservation of
the vector function
$\textbf{w}_{s}\equiv n_{s}\textbf{v}_{s\perp}
+\frac{q_{e}}{m_{e}c}\biggl(\Gamma_{s}-\frac{1}{c^{2}}\tilde{t}_{s}\biggr)\textbf{A}_{\perp}$,
while
vector function
$n_{s}v_{sz}\textbf{v}_{s\perp}
+\frac{q_{e}}{m_{e}c}(\Gamma_{s}v_{sz}+t_{sz})\textbf{A}_{\perp}$
gives the flux of function $\textbf{w}_{s}$.

In our analysis we assume the zero value of the conserving function
$\textbf{w}_{s}=0$.
It gives us the following expression for the transverse velocity $\textbf{v}_{s\perp}$ in terms of the vector potential $\textbf{A}_{\perp}$
\begin{equation}\label{RHDSEAW2022NLC v via A}\textbf{v}_{s\perp}
=-\frac{q_{e}}{m_{e}c}\biggl(\frac{\Gamma_{s}}{n_{s}}-\frac{\tilde{t}_{s}}{n_{s}c^{2}}\biggr)\textbf{A}_{\perp}.\end{equation}

Obtained expression for the transverse velocity allows to rewrite the equation for the transverse vector potential (\ref{RHDSEAW2022NLC eq for A z perp})
\begin{equation}\label{RHDSEAW2022NLC eq for A z perp 2}
\partial_{t}^{2}\textbf{A}_{\perp}-c^{2}\partial_{z}^{2} \textbf{A}_{\perp}
+\frac{\omega_{Ld}^{2}}{n_{0d}}
\biggl(\Gamma_{eu}+\Gamma_{ed}-\frac{1}{c^{2}}\tilde{t}_{u}-\frac{1}{c^{2}}\tilde{t}_{d}\biggr)\textbf{A}_{\perp}=0, \end{equation}
where $\omega_{Ld}^{2}=\frac{4\pi e^{2}n_{0d}}{m_{e}}$
is the partial Langmuir frequency for the spin-down electrons.

Expression (\ref{RHDSEAW2022NLC v via A}) allows to make transformation of equation (\ref{RHDSEAW2022NLC Euler Long z 1}).
However, equation (\ref{RHDSEAW2022NLC Euler Long z 1}) also contains the contribution of the transverse motion via the transverse part of the flux of the average reverse gamma factor $\textbf{t}_{s\perp}$.
Hence, we need to get relation between $\textbf{t}_{s\perp}$ and $\textbf{A}_{\perp}$,
which is found in Appendix A
\begin{equation}\label{RHDSEAW2022NLC} \textbf{t}_{s\perp}=\beta_{s} \textbf{A}_{\perp} \end{equation}
using equation (\ref{RHDSEAW2022NLC eq for t a}).
Explicit form of parameter $\beta_{s}$ is also given in Appendix A.

Let us to point out that considering the longitudinal motion
we consider the small amplitude acoustic waves.
Therefore, we can consider the linear response for the longitudinal perturbations.

This assumption allows to get further simplification of equation (\ref{RHDSEAW2022NLC eq for A z perp 2})
$$\partial_{t}^{2}\textbf{A}_{\perp}-c^{2}\partial_{z}^{2} \textbf{A}_{\perp}
+\frac{\omega_{Ld}^{2}}{n_{0d}}
\biggl(\Gamma_{0eu}-\frac{1}{c^{2}}\tilde{t}_{0u}$$
\begin{equation}\label{RHDSEAW2022NLC eq for A z perp 3}
+\Gamma_{0ed}-\frac{1}{c^{2}}\tilde{t}_{0d}
+\delta\Gamma_{eu}+\delta\Gamma_{ed}-\frac{1}{c^{2}}\delta\tilde{t}_{u}-\frac{1}{c^{2}}\delta\tilde{t}_{d}\biggr)\textbf{A}_{\perp}=0, \end{equation}
The equilibriums values of functions $\Gamma_{0s}$ and $\tilde{t}_{0s}$ can be found from the corresponding equations of state
(\ref{RHDSEAW2022NLC Gamma rel eq of state}) and (\ref{RHDSEAW2022NLC t rel eq of state}), correspondingly.
Next, the perturbations of $\delta\tilde{t}_{s}$ can be also found from equation (\ref{RHDSEAW2022NLC t rel eq of state}).
We can make expansion of expression (\ref{RHDSEAW2022NLC t rel eq of state}) on the perturbations of the concentration
and take the linear on $\delta n_{s}$ part.
Perturbations of relativistic gamma function $\delta\Gamma_{s}$ can be found from equation (\ref{RHDSEAW2022NLC Gamma z simpl}).
However, equation (\ref{RHDSEAW2022NLC Gamma z simpl}) requires the longitudinal part of the flux of the average reverse relativistic gamma factor $t_{sz}$.
Equation for $t_{sz}$ can be found from general equation (\ref{RHDSEAW2022NLC eq for t a}).

Before, we consider equation for $t_{sz}$ let us discuss equation for $v_{sz}$ and $n_{s}$.
Here, we ready to give representation of equation (\ref{RHDSEAW2022NLC Euler Long z 1}),
which is also considered in the linear regime on perturbations of functions $n_{s}$, $v_{sz}$, $\Gamma_{s}$ and $\varphi$
\begin{equation}\label{RHDSEAW2022NLC Euler Long z 2}
n_{0s}\partial_{t}v_{sz}+\frac{\delta p_{0s}}{\delta n_{0s}}\partial_{z}n_{s}
=-\frac{q_{e}}{m_{e}}\biggl(\Gamma_{0s}-\frac{1}{c^{2}}\tilde{t}_{0s}\biggr)\partial_{z}\varphi
+\Upsilon_{s}\partial_{z}\textbf{A}_{\perp}^{2}, \end{equation}
where
$n_{s}=n_{0s}+\delta n_{s}$, $\Gamma_{s}=\Gamma_{0s}+\delta\Gamma_{s}$,
$v_{sz}=\delta v_{sz}$ and $\varphi=\delta \varphi$,
and we also introduced parameter
$\Upsilon_{s}=-\frac{1}{2}(\frac{q_{e}}{m_{e}c})^{2}
[\Gamma_{0s}(\frac{\Gamma_{0s}}{n_{0s}}-\frac{\tilde{t}_{0s}}{n_{0s}c^{2}}) -\beta_{s}]$.
We take the derivative on $z$ of equation (\ref{RHDSEAW2022NLC Euler Long z 2}) and use the continuity equation (\ref{RHDSEAW2022NLC cont z})
$$\partial_{t}^{2}\delta n_{s}-\frac{\delta p_{0s}}{\delta n_{0s}}\partial_{z}^{2}\delta n_{s}$$
\begin{equation}\label{RHDSEAW2022NLC eq for n second order 1}
+\omega_{Ls}^{2}\biggl(\frac{\Gamma_{0s}}{n_{0s}}-\frac{\tilde{t}_{0s}}{n_{0s}c^{2}}\biggr)(\delta n_{u}+\delta n_{d})
-\Upsilon_{s}\partial_{z}^{2}\textbf{A}_{\perp}^{2}=0, \end{equation}
where we use the simplified  equation (\ref{RHDSEAW2022NLC eq for varphi gen})
\begin{equation}\label{RHDSEAW2022NLC eq for varphi z}
\partial_{z}^{2}\varphi=-4\pi q_{e}(\delta n_{eu}+\delta n_{ed}).\end{equation}

Let us present equation for $t_{sz}$ found from equation (\ref{RHDSEAW2022NLC eq for t a})
under assumption of the small amplitude response of the longitudinal functions,
and nonrelativistic fluxes $\textbf{v}^{2}\ll c^{2}$:
$$\partial_{t}\delta t_{sz}+\Gamma_{0s}\partial_{t}\delta v_{sz}
+\frac{\delta \tilde{t}_{0s}}{\delta n_{0s}}\partial\delta n_{s}$$
$$=-\frac{q_{e}}{m_{e}}n_{0s}\biggl(1-\frac{5p_{0s}}{n_{0s}c^{2}}+\frac{10M_{0s}}{3n_{0s}c^{4}}\biggr)\partial_{z}\varphi$$
\begin{equation}\label{RHDSEAW2022NLC eq t sz simpl}
-\frac{1}{2}\biggl(\frac{q_{e}}{m_{e}c}\biggr)^{2}n_{0s}\biggl(1-\frac{5p_{0s}}{n_{0s}c^{2}}\biggr)
\biggl(\frac{\Gamma_{0s}}{n_{0s}}-\frac{\tilde{t}_{0s}}{n_{0s}c^{2}}\biggr)
\partial_{z}\textbf{A}_{\perp}^{2}.\end{equation}

To exclude $\delta t_{sz}$ from final equations
we take the derivative on $z$ of equation (\ref{RHDSEAW2022NLC eq t sz simpl}) and use equation (\ref{RHDSEAW2022NLC Gamma z simpl})
$$\partial_{t}^{2}\delta\Gamma_{s}-\frac{\delta \tilde{t}_{0s}}{\delta n_{0s}}\partial_{z}^{2}\delta n_{s}
+\omega_{Ls}^{2}\biggl(1-\frac{5p_{0s}}{n_{0s}c^{2}}+\frac{10M_{0s}}{3n_{0s}c^{4}}\biggr)(\delta n_{u}+\delta n_{d})$$
\begin{equation}\label{RHDSEAW2022NLC eq Gamma second order 1} -\frac{1}{2}\biggl(\frac{q_{e}}{m_{e}c}\biggr)^{2}\biggl(1-\frac{5p_{0s}}{n_{0s}c^{2}}\biggr)\partial_{z}^{2}\textbf{A}_{\perp}^{2}=0. \end{equation}



Equations presented above are shown in the way as they appear from the general hydrodynamic equations
(\ref{RHDSEAW2022NLC cont via v}), (\ref{RHDSEAW2022NLC Euler for v}), (\ref{RHDSEAW2022NLC eq for Gamma}),
and (\ref{RHDSEAW2022NLC eq for t a}).
However, the application of equations of state (\ref{RHDSEAW2022NLC p rel eq of state}),
(\ref{RHDSEAW2022NLC t rel eq of state}), (\ref{RHDSEAW2022NLC M rel eq of state}), and
(\ref{RHDSEAW2022NLC Gamma rel eq of state}).
However, some combinations of parameters can be represented in physically more meaningful notations:
$(\frac{\Gamma_{0s}}{n_{0s}}-\frac{\tilde{t}_{0s}}{n_{0s}c^{2}})=\frac{1}{\gamma_{Fs}}$,
where
$\gamma_{Fe}=1/\sqrt{1-v_{Fe}^{2}/c^{2}}=\sqrt{1+p_{Fe}^{2}/m^{2}c^{2}}$,
and
$v_{Fe}^{2}=c^{2}\frac{p_{Fe}^{2}}{p_{Fe}^{2}+m^{2}c^{2}}
=\frac{p_{Fe}^{2}}{m^{2}\gamma_{Fe}^{2}}$,
with
$p_{Fe}=(3\pi^{2}n_{0e})^{1/3}\hbar$,
We also find
$(1-\frac{5p_{0s}}{n_{0s}c^{2}}+\frac{10M_{0s}}{3n_{0s}c^{4}})=\frac{1}{\gamma_{Fs}^{2}}$.

\section{The stability analysis}

Let us sum up the obtained wave equations in order to study the stability of the system.
We repeat equations (\ref{RHDSEAW2022NLC eq for A z perp 3}), (\ref{RHDSEAW2022NLC eq for n second order 1}),
(\ref{RHDSEAW2022NLC eq Gamma second order 1}) with some small modifications:
\begin{equation}\label{RHDSEAW2022NLC eq for n second order 2}
\partial_{t}^{2}\delta n_{s}-\frac{\delta p_{0s}}{\delta n_{0s}}\partial_{z}^{2}\delta n_{s}
+\frac{\omega_{Ls}^{2}}{\gamma_{Fs}}(\delta n_{u}+\delta n_{d})
-\Upsilon_{s}\partial_{z}^{2}\textbf{A}_{\perp}^{2}=0, \end{equation}
\begin{equation}\label{RHDSEAW2022NLC eq Gamma second order 2}
\partial_{t}^{2}\delta\Gamma_{s}-\frac{\delta \tilde{t}_{0s}}{\delta n_{0s}}\partial_{z}^{2}\delta n_{s}
+\frac{\omega_{Ls}^{2}}{\gamma_{Fs}^{2}}(\delta n_{u}+\delta n_{d})
-\Xi_{s}\partial_{z}^{2}\textbf{A}_{\perp}^{2}=0, \end{equation}
where
$\Xi_{s}=\frac{1}{2}(\frac{q_{e}}{m_{e}c})^{2}(1-\frac{5p_{0s}}{n_{0s}c^{2}})$,
and
$$\partial_{t}^{2}\textbf{A}_{\perp}-c^{2}\partial_{z}^{2} \textbf{A}_{\perp}
+\frac{\omega_{Ld}^{2}}{n_{0d}}
\biggl(\frac{n_{0u}}{\gamma_{Fu}}+\frac{n_{0d}}{\gamma_{Fd}}$$
\begin{equation}\label{RHDSEAW2022NLC eq for A z perp 4}
+\delta\Gamma_{eu}+\delta\Gamma_{ed}-\frac{1}{c^{2}}\delta\tilde{t}_{u}-\frac{1}{c^{2}}\delta\tilde{t}_{d}\biggr)\textbf{A}_{\perp}=0. \end{equation}

Start analysis of obtained equation from the transformation of equation (\ref{RHDSEAW2022NLC eq for A z perp 4}) explicitly using
structure (\ref{RHDSEAW2022NLC A structure}).
We also include $\partial_{t}A\ll\omega_{0}A$ since amplitude $A$ is the slow amplitude.
The linearized equation (\ref{RHDSEAW2022NLC eq for A z perp 4}) gives approximate relation between $\omega_{0}$ and $k_{0}$:
$\omega_{0}^{2}=k_{0}^{2}c^{2}+\frac{\omega_{Ld}^{2}}{\gamma_{Fd}}+\frac{\omega_{Lu}^{2}}{\gamma_{Fu}}$.
It is also used to cancel corresponding terms in equation (\ref{RHDSEAW2022NLC eq for A z perp 4}):
$$2\imath\omega_{0}(\partial_{t}A+ V_{g}\partial_{z}A)+U^{2}\partial_{z}^{2}A$$
\begin{equation}\label{RHDSEAW2022NLC eq for A z perp 4}
-\frac{\omega_{Ld}^{2}}{n_{0d}}
\biggl(\delta\Gamma_{eu}+\delta\Gamma_{ed}
-\frac{1}{c^{2}}\delta\tilde{t}_{u}-\frac{1}{c^{2}}\delta\tilde{t}_{d}\biggr)A=0,
\end{equation}
where $V_{g}=k_{0}c^{2}/\omega_{0}$, and $U^{2}=c^{2}$.


Function $A(z,t)$ in equation (\ref{RHDSEAW2022NLC eq for A z perp 4}) is the slowly changing complex amplitude.
We split it on the amplitude and the phase $A(z,t)=a(z,t)e^{\imath\theta(z,t)}$,
with further decomposition of the amplitude and the phase
$a(z,t)=\frac{1}{2}[a_{0}+\delta a \cdot e^{\imath kz-\imath\Omega t}+c.c.]$,
$\delta A(z,t)=\frac{1}{2}[\delta a \cdot e^{\imath kz-\imath\Omega t}+c.c.]$
$\theta(z,t)=\frac{1}{2}[\theta_{0}+\delta \theta \cdot e^{\imath kz-\imath\Omega t}+c.c.]$,
where
$\delta a\ll a_{0}$,
and
$\delta \theta\ll \theta_{0}$.
Next, we make corresponding decomposition of the hydrodynamic functions
$\delta n_{s}=\frac{1}{2}[N_{s}\cdot e^{\imath kz-\imath\Omega t}+c.c.]$,
and
$\delta \Gamma_{s}=\frac{1}{2}[G_{s}\cdot e^{\imath kz-\imath\Omega t}+c.c.]$.

\begin{figure} \includegraphics[width=8cm,angle=0]{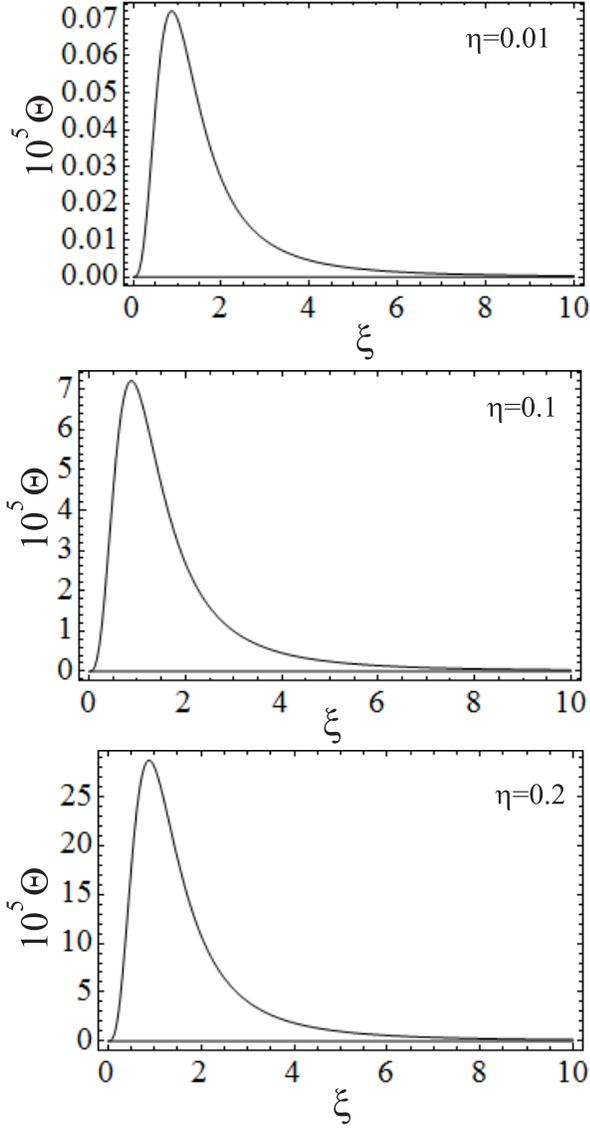}
\caption{\label{RHD2021ClLM Fig 01}
Increment of instability is proportional to parameter $\Theta$ (\ref{RHDSEAW2022NLC DISP eq 4}).
Parameter $\Theta$ is shown as the function of dimensionless Fermi momentum $\xi$
which is proportional to the concentration of electrons.
This dependence is made for several values of the spin polarization $\eta$ of plasmas.
}\end{figure}

We use presented decompositions in order to simplify equations
(\ref{RHDSEAW2022NLC eq for n second order 2}), (\ref{RHDSEAW2022NLC eq Gamma second order 2}),
and (\ref{RHDSEAW2022NLC eq for A z perp 4})
\begin{equation}\label{RHDSEAW2022NLC eq for n second order 3}
-\Omega^{2}\delta n_{s}+\frac{\delta p_{0s}}{\delta n_{0s}}k^{2}\delta n_{s}
+\frac{\omega_{Ls}^{2}}{\gamma_{Fs}}(\delta n_{u}+\delta n_{d})
+\Upsilon_{s}k^{2}a_{0}\delta A=0, \end{equation}
and
\begin{equation}\label{RHDSEAW2022NLC eq Gamma second order 3}
-\Omega^{2}\delta\Gamma_{s}+\frac{\delta \tilde{t}_{0s}}{\delta n_{0s}}k^{2}\delta n_{s}
+\frac{\omega_{Ls}^{2}}{\gamma_{Fs}^{2}}(\delta n_{u}+\delta n_{d})
+\Xi_{s}k^{2}a_{0}\delta A=0, \end{equation}
while equation (\ref{RHDSEAW2022NLC eq for A z perp 4}) splits on two equations
for the amplitude and the phase
\begin{equation}\label{RHDSEAW2022NLC eq for a}
2\omega_{0}\partial_{t}a+2\omega_{0}V_{g}\partial_{z}a+U^{2}a_{0}\partial_{z}^{2}\theta=0,\end{equation}
and
\begin{equation}\label{RHDSEAW2022NLC eq for theta}
2\omega_{0}a_{0}\partial_{t}\theta+ 2\omega_{0}V_{g}a_{0}\partial_{z}\theta
-U^{2}\partial_{x}^{2}a+a_{0}\delta H =0, \end{equation}
where
\begin{equation}\label{RHDSEAW2022NLC H def} \delta H=\frac{\omega_{Ld}^{2}}{n_{0d}}
\biggl(\delta\Gamma_{eu}+\delta\Gamma_{ed}
-\frac{1}{c^{2}}\delta\tilde{t}_{u}-\frac{1}{c^{2}}\delta\tilde{t}_{d}\biggr). \end{equation}

Equations (\ref{RHDSEAW2022NLC eq for a}) and (\ref{RHDSEAW2022NLC eq for theta}) gives
\begin{equation}\label{RHDSEAW2022NLC eq for delta a}
\biggl[(\Omega-V_{g}k)^{2}-\frac{1}{4}\frac{U^{4}k^{4}}{\omega_{0}^{2}}\biggr]\delta a -a_{0}\delta H=0. \end{equation}

We get expressions for the partial concentrations from equation (\ref{RHDSEAW2022NLC eq for n second order 3})
\begin{equation}\label{RHDSEAW2022NLC n u via A}
\delta n_{u}=\frac{a_{0}k^{2}\delta A}{\Delta_{N}}
\biggl[\Upsilon_{u}\biggl(\Omega^{2}-u_{pd}^{2}k^{2}-\frac{\omega_{Ld}^{2}}{\gamma_{Fd}}\biggr)
+\Upsilon_{d}\frac{\omega_{Lu}^{2}}{\gamma_{Fu}}\biggr], \end{equation}
and
\begin{equation}\label{RHDSEAW2022NLC n d via A}
\delta n_{d}=\frac{a_{0}k^{2}\delta A}{\Delta_{N}}
\biggl[\frac{\omega_{Ld}^{2}}{\gamma_{Fd}}\Upsilon_{u}
+\Upsilon_{d}\biggl(\Omega^{2}-u_{pu}^{2}k^{2}-\frac{\omega_{Lu}^{2}}{\gamma_{Fu}}\biggr)\biggr],\end{equation}
where
$$\Delta_{N}= (\Omega^{2}-u_{pd}^{2}k^{2})(\Omega^{2}-u_{pu}^{2}k^{2})$$
\begin{equation}\label{RHDSEAW2022NLC Delat N} -\frac{\omega_{Lu}^{2}}{\gamma_{Fu}}(\Omega^{2}-u_{pd}^{2}k^{2})
-\frac{\omega_{Ld}^{2}}{\gamma_{Fd}}(\Omega^{2}-u_{pu}^{2}k^{2}), \end{equation}
with $u_{ps}^{2}=\frac{\delta p_{0s}}{\delta n_{0s}}$.

We need to get the contribution of the partial hydrodynamic gamma functions $\Gamma_{s}$.
However, we do not need expressions for each of them.
We need to find their sum $\delta\Gamma_{u}+\delta\Gamma_{d}$ only,
as we can see it from expression for $\delta H$ (see equation (\ref{RHDSEAW2022NLC H def})).
We find required expression from equation (\ref{RHDSEAW2022NLC eq Gamma second order 3})
$$\delta \Gamma_{u}+\delta \Gamma_{d}=\frac{1}{\Omega^{2}}
\biggl[\frac{\delta \tilde{t}_{0u}}{\delta n_{0u}}k^{2}\delta n_{u}+\frac{\delta \tilde{t}_{0d}}{\delta n_{0d}}k^{2}\delta n_{d}$$
\begin{equation}\label{RHDSEAW2022NLC Gamma ud via A}
+(\Xi_{u}+\Xi_{d})a_{0}k^{2}\delta A
+\biggl(\frac{\omega_{Lu}^{2}}{\gamma_{Fu}^{2}}+\frac{\omega_{Ld}^{2}}{\gamma_{Fd}^{2}}\biggr)(\delta n_{u}+\delta n_{d})\biggr]. \end{equation}
We also need $\delta n_{u}+\delta n_{d}$:
\begin{equation}\label{RHDSEAW2022NLC}
\delta n_{u}+\delta n_{d}=\frac{a_{0}k^{2}\delta A}{\Delta_{N}}
[\Upsilon_{u}(\Omega^{2}-u_{pd}^{2}k^{2})+\Upsilon_{d}(\Omega^{2}-u_{pu}^{2}k^{2})]. \end{equation}


Obtained expressions for the partial concentrations $\delta n_{s}$ and gamma functions $\delta\Gamma_{s}$
allow us to represent function $\delta H$ (\ref{RHDSEAW2022NLC H def})
$$\delta H=\frac{\omega_{Ld}^{2}}{n_{0d}} \frac{1}{\Omega^{2}c^{2}}\biggl[
\biggl(\frac{\delta \tilde{t}_{0u}}{\delta n_{0u}}\delta n_{u}+\frac{\delta \tilde{t}_{0d}}{\delta n_{0d}}\delta n_{d}\biggr)
(k^{2}c^{2}-\Omega^{2})$$
\begin{equation}\label{RHDSEAW2022NLC} +c^{2}(\Xi_{u}+\Xi_{d})a_{0}k^{2}\delta A
+c^{2}\biggl(\frac{\omega_{Lu}^{2}}{\gamma_{Fu}^{2}}+\frac{\omega_{Ld}^{2}}{\gamma_{Fd}^{2}}\biggr)(\delta n_{u}+\delta n_{d})\biggr]. \end{equation}
Moreover, we get the following consequence of equation (\ref{RHDSEAW2022NLC eq for delta a}):
$$\Delta_{N}\cdot[(\Omega-V_{g}k)^{2}-\frac{1}{4}\frac{U^{4}k^{4}}{\omega_{0}^{2}}]
=-\frac{1}{2}\frac{q_{e}^{2}}{m_{e}^{2}c^{2}}\frac{a_{0}^{2}k^{2}}{\Omega^{2}c^{2}}\frac{\omega_{Ld}^{2}}{n_{0d}}\times$$
$$\times \biggl\{\frac{\delta \tilde{t}_{0d}}{\delta n_{0d}}(k^{2}c^{2}-\Omega^{2})
\biggl[ \frac{n_{0d}\omega_{Lu}^{2}}{\gamma_{Fd}\gamma_{Fu}^{2}}
+\frac{n_{0d}}{\gamma_{Fd}^{2}}\biggl(\Omega^{2}-U_{pu}^{2}k^{2}-\frac{\omega_{Lu}^{2}}{\gamma_{Fu}}\biggr)\biggr]$$
$$+\frac{\delta \tilde{t}_{0u}}{\delta n_{0u}} (k^{2}c^{2}-\Omega^{2})
\biggl[\frac{n_{0u}}{\gamma_{Fu}^{2}}\biggl(\Omega^{2}-U_{pd}^{2}k^{2}-\frac{\omega_{Ld}^{2}}{\gamma_{Fd}}\biggr)
+\frac{n_{0u}\omega_{Ld}^{2}}{\gamma_{Fu}\gamma_{Fd}^{2}}\biggr]$$
\begin{equation}\label{RHDSEAW2022NLC DISP eq 1}
+c^{2}\biggl(\frac{\omega_{Lu}^{2}}{\gamma_{Fu}^{2}} +\frac{\omega_{Ld}^{2}}{\gamma_{Fd}^{2}}\biggr)
\biggl[ \frac{n_{0u}}{\gamma_{Fu}^{2}}(\Omega^{2}-U_{pd}^{2}k^{2}) +\frac{n_{0d}}{\gamma_{Fd}^{2}}(\Omega^{2}-U_{pu}^{2}k^{2})\biggr]
\biggr\},\end{equation}
where
$(\Xi_{u}+\Xi_{d})\cdot\Delta_{N}\approx0$.

The high-frequency regime of the matter waves with the spectrum obtained from $\Delta_{N}\approx0$
consists of two waves: the Langmuir wave and the spin-electron-acoustic wave.
Depending on the equilibrium concentration of electrons $n_{0e}=n_{0u}+n_{0d}$
the frequency of these waves can be comparable or they can have considerable difference.
For the large concentrations $n_{0e}^{1/3}\hbar/m_{e}c\sim1$ we have comparable frequencies of these waves in the high-frequency regime,
which also corresponds to relatively large wave vectors.

Let us consider the low-frequency regime,
where we have single spin-electron-acoustic wave.
This regime of relatively small frequencies corresponds to frequencies
high enough to neglect the motion of ions and the contribution of the ion-acoustic wave.
The chosen small frequency regime corresponds to the small wave vectors,
where $\Omega_{SEAW}=c_{A}k\ll\frac{\omega_{Ls}^{2}}{\gamma_{Fs}}$,
and $c_{A}$ has same order as $U_{ps}$.
It also gives restriction on the large spin polarizations.

We make simplification of $\Delta_{N}$ and the right-hand side of equation (\ref{RHDSEAW2022NLC DISP eq 1})
in the chosen range of parameters.
First, we consider parameter $\Delta_{N}$ (\ref{RHDSEAW2022NLC Delat N}).
We present it as the superposition of two parts $\Delta_{N}=\Delta_{N1}+\Delta_{N2}$,
where
$\Delta_{N1}
=\Omega^{2}[\Omega^{2}- (u_{pd}^{2}k^{2} +u_{pu}^{2}k^{2} +\omega_{Lu}^{2}/\gamma_{Fu} +\omega_{Ld}^{2}/\gamma_{Fd})
\approx -\Omega^{2}(\omega_{Lu}^{2}/\gamma_{Fu}+\omega_{Ld}^{2}/\gamma_{Fd})$,
and
$\Delta_{N2}=u_{pd}^{2}u_{pu}^{2}k^{4}+ u_{pd}^{2}k^{2} \omega_{Lu}^{2}/\gamma_{Fu} +u_{pu}^{2}k^{2} \omega_{Ld}^{2}/\gamma_{Fd}
\approx u_{pd}^{2}k^{2} \omega_{Lu}^{2}/\gamma_{Fu} +u_{pu}^{2}k^{2} \omega_{Ld}^{2}/\gamma_{Fd}$.
Hence, we find
\begin{equation}\label{RHDSEAW2022NLC Delat N 2}
\Delta_{N}\approx -\frac{\omega_{Lu}^{2}}{\gamma_{Fu}}(\Omega^{2}-u_{pd}^{2}k^{2})
-\frac{\omega_{Ld}^{2}}{\gamma_{Fd}}(\Omega^{2}-u_{pu}^{2}k^{2}). \end{equation}
Condition $\Delta_{N}=0$ gives the following spectrum of acoustic waves $\Omega=c_{A}k$,
with
\begin{equation}\label{RHDSEAW2022NLC Spectrum SEAW} c_{A}^{2}
=\frac{u_{pd}^{2}\frac{\omega_{Lu}^{2}}{\gamma_{Fu}}
+u_{pu}^{2}\frac{\omega_{Ld}^{2}}{\gamma_{Fd}}}{\frac{\omega_{Lu}^{2}}{\gamma_{Fu}}+\frac{\omega_{Ld}^{2}}{\gamma_{Fd}}}, \end{equation}
where $u_{ps}^{2}=(c^{2}/3)p_{Fs}^{2}/(p_{Fs}^{2}+m_{e}^{2}c^2)$.

Using suggested range of parameters and spectrum of spin-electron-acoustic wave
we find simplified form of equation (\ref{RHDSEAW2022NLC DISP eq 1})
$$(\Omega^{2}-c_{A}^{2}k^{2})\cdot[(\Omega-V_{g}k)^{2}-\frac{1}{4}\frac{U^{4}k^{4}}{\omega_{0}^{2}}]
=\frac{1}{2}\frac{q_{e}^{2}}{m_{e}^{2}c^{2}}\frac{a_{0}^{2}k^{2}}{\Omega^{2}c^{2}}\times$$
$$\times\frac{\omega_{Lu}^{2}\omega_{Ld}^{2}}{\gamma_{Fu}\gamma_{Fd}}
\frac{\frac{1}{\gamma_{Fu}}-\frac{1}{\gamma_{Fd}}}{(\frac{\omega_{Lu}^{2}}{\gamma_{Fu}} +\frac{\omega_{Ld}^{2}}{\gamma_{Fd}})}
\biggl\{ (k^{2}c^{2}-\Omega^{2})
\biggl(\frac{\delta \tilde{t}_{0d}}{\delta n_{0d}}-\frac{\delta \tilde{t}_{0u}}{\delta n_{0u}}\biggr) $$
\begin{equation}\label{RHDSEAW2022NLC DISP eq 2}
-k^{2}c^{2}(U_{pd}^{2}-U_{pu}^{2})
\frac{\biggl(\frac{\omega_{Lu}^{2}}{\gamma_{Fu}^{2}}
+\frac{\omega_{Ld}^{2}}{\gamma_{Fd}^{2}}\biggr)}{\frac{\omega_{Lu}^{2}}{\gamma_{Fu}} +\frac{\omega_{Ld}^{2}}{\gamma_{Fd}}}
\biggr\},\end{equation}
where we used
$\Delta_{N}=-(\Omega^{2}-c_{A}^{2}k^{2})(\frac{\omega_{Lu}^{2}}{\gamma_{Fu}} +\frac{\omega_{Ld}^{2}}{\gamma_{Fd}})$.

In order to consider the instability of the system
we assume
$\Omega=V_{g}k+\frac{1}{2}\frac{U^{2}k^{2}}{\omega_{0}}+\imath \Lambda$
and
$\Omega=c_{A}k+\imath\Lambda$.
The low frequency branch of the electromagnetic wave can be chosen
$\Omega_{-}=V_{g}k-\frac{1}{2}\frac{U^{2}k^{2}}{\omega_{0}}$,
but it appears to be stable in the considered regime.

It leads to
$$\Lambda^{2}=\frac{-1}{2c_{A}k}\frac{\omega_{0}}{U^{2}k^{2}}
\frac{1}{2}\frac{q_{e}^{2}}{m_{e}^{2}c^{2}}\frac{a_{0}^{2}k^{2}}{\Omega^{2}c^{2}}\times$$
$$\times\frac{\omega_{Lu}^{2}\omega_{Ld}^{2}}{\gamma_{Fu}\gamma_{Fd}}
\frac{\frac{1}{\gamma_{Fu}}-\frac{1}{\gamma_{Fd}}}{(\frac{\omega_{Lu}^{2}}{\gamma_{Fu}} +\frac{\omega_{Ld}^{2}}{\gamma_{Fd}})}
\biggl\{ (k^{2}c^{2}-\Omega^{2})
\biggl(\frac{\delta \tilde{t}_{0d}}{\delta n_{0d}}-\frac{\delta \tilde{t}_{0u}}{\delta n_{0u}}\biggr) $$
\begin{equation}\label{RHDSEAW2022NLC DISP eq 3}
-k^{2}c^{2}(U_{pd}^{2}-U_{pu}^{2})
\frac{\biggl(\frac{\omega_{Lu}^{2}}{\gamma_{Fu}^{2}}
+\frac{\omega_{Ld}^{2}}{\gamma_{Fd}^{2}}\biggr)}{\frac{\omega_{Lu}^{2}}{\gamma_{Fu}} +\frac{\omega_{Ld}^{2}}{\gamma_{Fd}}}
\biggr\}.\end{equation}

Next, let us represent it in the dimensionless form
\begin{equation}\label{RHDSEAW2022NLC DISP eq 4}
\frac{\Lambda^{2}}{\omega_{Le}^{2}}=\frac{\omega_{0}}{\omega_{Le}}
\frac{q_{e}^{2}a_{0}^{2}}{m_{e}^{2}U^{2}c^{2}}\cdot\Theta, \end{equation}
where
$$\Theta=-\frac{1}{8}\frac{1}{\frac{c_{A}^{3}k^{3}}{\omega_{Le}^{3}}}\frac{1-\eta^{2}}{\gamma_{Fu}\gamma_{Fd}}
\frac{\frac{1}{\gamma_{Fu}}-\frac{1}{\gamma_{Fd}}}{(\frac{1-\eta}{\gamma_{Fu}} +\frac{1+\eta}{\gamma_{Fd}})}
\biggl[\frac{(k^{2}c^{2}-\Omega^{2})}{\omega_{Le}^{2}}
\biggl(\frac{u_{td}^{2}}{c^{2}}-\frac{u_{tu}^{2}}{c^{2}}\biggr)$$
\begin{equation}\label{RHDSEAW2022NLC Theta for Lambda}
-\frac{k^{2}c^{2}}{\omega_{Le}^{2}}\biggl(\frac{U_{pd}^{2}}{c^{2}}-\frac{U_{pu}^{2}}{c^{2}}\biggr)
\frac{(\frac{1-\eta}{\gamma_{Fu}^{2}} +\frac{1+\eta}{\gamma_{Fd}^{2}})}{(\frac{1-\eta}{\gamma_{Fu}} +\frac{1+\eta}{\gamma_{Fd}})}
\biggr],\end{equation}
where
$\eta=\mid n_{0d}-n_{0u}\mid/(n_{0d}+n_{0u})$,
and
$u_{ts}^{2}=\frac{\delta \tilde{t}_{0s}}{\delta n_{0s}}$,
with
$u_{ts}^{2}=(c^{2}/3)p_{Fs}^{2}/(p_{Fs}^{2}+m_{e}^{2}c^2)^{3/2}$.

Numerical analysis of the increment of instability $\Lambda$ is made in Fig. \ref{RHD2021ClLM Fig 01},
where it is demonstrated via study of behavior of $\Theta$ (\ref{RHDSEAW2022NLC Theta for Lambda}).
We see that the instability exists in the area of intermediate concentrations,
which however correspond to the interval of relatively large concentrations $n_{0e}\sim 10^{30}$ cm$^{-3}$.
Strong increase of function $\Theta$ with the increase of the spin polarization is also can be seen in Fig. \ref{RHD2021ClLM Fig 01}.



\section{Conclusion}

Propagation of strong electromagnetic waves through the high density degenerate plasmas has been considered.
It has been considered in order to study the radiation of compact astrophysical objects,
which propagates through the layer of plasmas,
after the generation of radiation.
Particularly, we have studied effect induced by the interaction of electromagnetic waves with plasmas, particularly, with the small frequency acoustic waves.
Moreover, it has been assumed that the propagating radiation induced the spin polarization of plasmas.
Hence, the conditions for existence of the spin-electron-acoustic waves has been created.
We have considered the two-component electron-ion plasmas with the assumption of motionless ions.
But, the electrons being spin polarized can be considered as two fluids with different spin projections.
Hence, we have two active fluids interaction with the radiation.
For relatively small spin polarization both subspecies of electrons are degenerate at the same conditions.

The strong
nonlinear coupling between the electromagnetic waves and
spin-electron-acoustic waves leads to decay instability.
This instability exists in the interval of concentrations of electrons
near $p_{Fe}=\sqrt[3]{3\pi^{2}n_{0e}}\hbar \approx m_{e}c$
and depends on the spin polarization of electrons showing strong increase with the increase of the spin polarization.
Basically it exists in the nonrelativistic limit,
but has small value.
The instability also disappears at large concentrations corresponding to ultrarelativistic Fermi momentums.


\section{DATA AVAILABILITY}

Data sharing is not applicable to this article as no new data were
created or analyzed in this study, which is a purely theoretical one.

\appendix

\section{Calculation of transverse flux of the average relativistic gamma factor}


Function $\textbf{t}_{s\perp}$ enters equation (\ref{RHDSEAW2022NLC Euler Long z 1}) in front of $\partial_{z}\textbf{A}_{\perp}$.
Equation of (\ref{RHDSEAW2022NLC Euler Long z 1}) describes the longitudinal dynamics,
which can be considered in the linear approximation.
Therefore, we can find expression for $\textbf{t}_{s\perp}$ in the linear approximation as well.
We can present the linearized equation for $\textbf{t}_{s\perp}$ from general equation (\ref{RHDSEAW2022NLC eq for t a}):
$$\partial_{t}\textbf{t}_{s\perp}+\Gamma_{0s}\partial_{t}\textbf{v}_{s\perp}$$
\begin{equation}\label{RHDSEAW2022NLC} =-\frac{q_{s}}{m_{s}c}n_{0s}\partial_{t}\textbf{A}_{\perp}
\biggl(1-\frac{5p_{0s}}{n_{0s}c^{2}}+\frac{10M_{0s}}{3n_{0s}c^{4}}\biggr).\end{equation}
It can be considered as the time derivative of one vector function.
Hence, this function is the constant.
Similarly to condition $\textbf{w}_{s}=0$
we assume that this function is also equal to zero.
We also include the expression for $\textbf{v}_{s\perp}$ via $\textbf{A}_{\perp}$ (\ref{RHDSEAW2022NLC v via A}).
Finally we find $\textbf{t}_{s\perp}=\beta_{s}\textbf{A}_{\perp}$,
where
\begin{equation}\label{RHDSEAW2022NLC beta s}
\beta_{s}= \frac{q_{s}}{m_{s}c}\biggl[\Gamma_{0s}\biggl(\frac{\Gamma_{0s}}{n_{0s}}-\frac{\tilde{t}_{0s}}{n_{0s}c^{2}}\biggr)
-\biggl(1-\frac{5p_{0s}}{n_{0s}c^{2}}+\frac{10M_{0s}}{3n_{0s}c^{4}}\biggr)\biggr].
\end{equation}


\end{document}